\documentclass[11pt,a4paper]{article}

\pdfoutput=1
\usepackage{jheppub}

\definecolor{rust}{rgb}{0.8,0.2,0.2}
\definecolor{green}{rgb}{0.1,0.8,0.2}

\newcommand{\de}{\partial}
\newcommand{\be}{\begin{equation}}
\newcommand{\ba}{\begin{eqnarray}}
\newcommand{\ea}{\end{eqnarray}}
\newcommand{\ee}{\end{equation}}

\newcommand{\f}{\frac}
\newcommand{\s}{\sqrt}

\newcommand{\ti}{\tilde}
\newcommand{\ap}{\alpha}

\newcommand{\ddd}{\cdot\cdot\cdot}
\newcommand{\no}{\nonumber \\}
\newcommand{\la}{\langle}
\newcommand{\lb}{\rangle}

\title{Entropic Counterpart of Perturbative Einstein Equation}

\author[a]{Jyotirmoy Bhattacharya}
\author[b,a]{and Tadashi Takayanagi}
\affiliation[a]{Kavli Institute for the Physics and Mathematics of the Universe (WPI),\\
 The University of Tokyo, Kashiwa, Chiba 277-8583, Japan.}
\affiliation[b]{Yukawa Institute for Theoretical Physics,
Kyoto University, Kyoto 606-8502, Japan.}

\emailAdd{jyotirmoy.bhattacharya@ipmu.jp}
\emailAdd{takayana@yukawa.kyoto-u.ac.jp}

\abstract{
Entanglement entropy in a field theory, with a holographic dual,
may be viewed as a quantity  which encodes the diffeomorphism invariant
bulk gravity dynamics. This, in particular, indicates that the bulk Einstein equations
would imply some constraints for the boundary entanglement entropy.
In this paper we focus on the change in entanglement entropy, for small but arbitrary
fluctuations about a given state, and analyze the constraints
imposed on it by the perturbative Einstein equations,
linearized about the corresponding bulk state. Specifically, we consider linear
fluctuations about BTZ black hole in 3 dimension, pure AdS and
AdS Schwarzschild black holes in 4 dimensions and obtain a diffeomorphism
invariant reformulation of linearized Einstein equation in terms of
holographic entanglement entropy. We will also show that
entanglement entropy for boosted subsystems provides the information 
about all the components of the metric with a time index.
}

\begin{document}

\begin{flushright} 
\small{YITP-13-69} \\
\small{IPMU13-0157} 
\end{flushright}

\maketitle

\flushbottom


\section{Introduction}

The principle of holography \cite{Hol}, especially the AdS/CFT correspondence \cite{Maldacena}, provides us
with a remarkable tool to study quantum gravity on a spacetime with boundaries. In the
bulk gravity theory, due to diffeomorphism invariance the choice of a coordinate system
is arbitrary and ambiguous, which sometimes makes the physical characterization of
local quantities difficult. On the other hand, in the boundary, where
the holographic dual lives, the metric is no longer dynamical and
we can choose a particular natural coordinate system (e.g. a flat spacetime), in the
boundary. Therefore, we can study the bulk gravity dynamics by using this fixed coordinate
on the boundary. If we take into account large quantum gravity effects,
we expect that the metric fluctuates a lot and will not be a good quantity
to characterize the geometry of spacetime, in the bulk. Nevertheless, as has often
been emphasized in AdS/CFT, these quantum effects can be non-perturbatively
well described in its holographic theory at the boundary. For these reasons, it is
reasonable to search for an appropriate quantity in the boundary field theory,
which encodes the diffeomorphism invariant information of the bulk dynamics.
Entanglement entropy in the boundary field theory, is a prominent candidate for such a
quantity.

The entanglement entropy (EE) provides us with a desirable universal quantity which
connects the gravity and its holographic dual (for reviews of EE
refer to e.g.\cite{Ereview,CCreview,CHreview,Lreview,HEEreview}). It can be defined in any quantum many-body
systems and includes quite a lot of information as we can choose infinitely many
different subsystems where we trace out the density matrix. The holographic
entanglement entropy (HEE) \cite{RT,HRT,HEEreview} tells us that in the gravity dual, assuming classical limit,
the EE is equal to the area of extremal surface in the bulk, which is manifestly diffeomorphism invariant. Recently,
this has been derived from the bulk to boundary relation \cite{GKPW} in \cite{LM} by
improving the argument in \cite{Fu} (see also \cite{CHM,He,Ha,Fa,FPS}
for other progresses in this direction). In the presence of quantum corrections \cite{Har,FLM} and
higher derivatives \cite{MyH,BKP,FPS} (see also \cite{CZ,BKS}), we need to modify this area law formula,
which can still be called holographic entanglement entropy (HEE). Even though the metric may not be a good
quantity which characterizes the spacetime in quantum gravity, the HEE is robust against quantum corrections
and thus is one of the candidates of fundamental variables in quantum gravity.

In the classical gravity limit, at least heuristically, we expect to reproduce the bulk metric of a given gravitational spacetime
once we know the holographic entanglement entropy for {\it any} choice of subsystem in the boundary, even without
the knowledge of the bulk Einstein equation
\footnote{We must emphasize that for obtaining the entire metric, the information of entanglement entropy for
a single type of subsystem with a given shape (like the round ball for all values of its radius) may not be enough.
This is simply because the metric has several components. So in order to ensure that
the entire metric is recoverable one may specify EE for different shapes of subsystems as well as EE for boosted subsystems. 
The latter, for example, is needed to get all the components of the metric
with a time index. In the AdS$_4$ example we will discuss in this paper, 
we just need the EE for round spheres with arbitrary radius and its boosted one in order to fully recover the 
perturbed metric once we impose the Einstein equation.}
(refer to \cite{met} for related discussions). In this correspondence,
roughly, the size of the subsystem is interpreted as the coordinate of extra dimension of the gravity dual.
Since Einstein equations constrain the space of all possible metrics, so it is also
expected to impose constraints on the holographic entanglement entropy.
In order to understand how EE encodes the metric, it is natural to ask what kind of constraints
for EE can be obtained from the bulk Einstein equation, for various
choices of the matter energy-momentum tensor and cosmological constant. In other words,
it is intriguing to ask what is a counterpart of Einstein equation for boundary EE. The paper \cite{NNPT}
made a first step in this direction and this question was studied mainly in pure AdS$_3$ within
perturbative calculations.

In this paper, we extend this analysis to rewrite perturbative Einstein equation in terms of
HEE in higher dimensional examples of AdS/CFT as well as the AdS black holes.
We primarily focus on the change in HEE corresponding to
linearized fluctuations about a given bulk state. Although we consider
the background state to be static, we make no symmetry assumptions
about the fluctuations. Even though in the AdS$_3/$CFT$_2$, the authors of \cite{NNPT}
found that the Einstein equation gives strong constraint which
determine the time evolution of HEE. In the present paper, we  find that the
Einstein equation itself is not that strong in AdS$_{d+1}/$CFT$_{d}$ ($d\geq 3$), but
it leads to a constraint equation which the HEE should satisfy at any time. We will try to
give a plausible interpretation of this constraint. We will also derive the constraint equation
for EE, in case of dual AdS black holes (in the limit, where the subsystem size is large)
and find an intriguing connection between the behavior of the HEE and that of
the quasi normal modes in AdS black holes.

This paper is organized as follows: In \S \ref{ssec:gs} we describe the general strategy that
we shall adopt in this paper. In \S \ref{sec:BTZ}, we will study the HEE in BTZ black holes and 
find the counterpart of Einstein equation in this case. In \S \ref{sec:AdS4} , we will analyze a
similar problem for the AdS$_4$ space. In \S \ref{sec:AdS4BH}, we generalize our argument to the
AdS$_4$ black hole example. In \S \ref{sec:boost}, we consider the HEE for boosted subsystems to
obtain a complete basis to cover all components of the metric.

\subsection{General Strategy} \label{ssec:gs}

We consider a  $d+1$  dimensional static space time of the form
\begin{equation}\label{staticmet}
\begin{split}
 ds^2 &=\f{R^2}{z^2}\left(\f{dz^2}{f(z)}+\tilde \eta_{\mu\nu}(z,x)dx^\mu dx^\nu \right). \\
 \end{split}
\end{equation}
where $\tilde \eta_{\mu\nu}$ is the diagonal matrix with the diagonal entries $\{-f(z), 1, 1 \dots \}$.
Note that when $f(z) = 1$ this metric \eqref{staticmet} is simply $AdS_{d+1}$ and $\tilde \eta$ is simply the Minkowski metric.
Also we consider the case, when $f(z) = 1 - m z^{d}$, the metric \eqref{staticmet} then represents the
Schwarzschild black hole in $AdS_{d+1}$. In the $z \rightarrow 0$ limit, the coordinate
$x^\mu$ ($\mu=0,1,2,\ddd,d-1$) describes the $d$ dimensional Lorentzian space $R^{1,d-1}$
where the $d$ dimensional conformal field theory (CFT$_d$) lives. The parameter $R$ describes the
radius of the AdS space.

We then consider linear perturbations around \eqref{staticmet} in the Fefferman-Graham (FG) gauge so that
the complete metric now takes the form
\be ds^2 =\f{R^2}{z^2}\left(\f{dz^2}{f(z)}+g_{\mu\nu}(z,x)dx^\mu dx^\nu \right). \label{adsmet} \ee

We now express the metric perturbation as follows:
 \be g_{\mu\nu}=\tilde \eta_{\mu\nu}+h_{\mu\nu}, \label{pert} \ee
 assuming that $h_{\mu\nu}$ is infinitesimally small,
 we will keep only the linear order of $h_{\mu\nu}$.
 This perturbation satisfies the Einstein equation as usual:
\be
{\cal R}_{ab}-\f{1}{2}{\cal R}g_{ab}-\f{d(d-1)}{2R^2}g_{ab}=T^{(G)}_{ab},
\label{Ein}
\ee
 where $T^{(G)}_{ab}$ is the energy stress tensor for
 the matter coupled to the Einstein gravity ($a,b=0,\ddd,d$ are indices of coordinates of the $d+1$ dimensional space).

The holographic entanglement entropy (HEE) in a $d+1$ dimensional AdS
gravity is given in terms of the area of $d-1$ dimensional extremal surface $\gamma_A$ as
\be
S_A=\f{\mbox{Area}(\gamma_A)}{4G_N}, \label{HEEarea}
\ee
so that the boundary of $\gamma_A$ coincides with that of the subsystem $A$ \cite{RT,HRT}.

In this paper we focus on the first order correction of HEE under the metric perturbation about the static
background geometry \eqref{staticmet}. This first order contribution
can be systematically computed as follows. First we consider the extremal surface $\gamma_A$ whose shape
is already known in the background geometry and calculate its area.
Since the background spacetime is static, $\gamma_A$ is a minimal area surface on the canonical time slice.
Next we evaluate the area of the same surface $\gamma_A$ in the perturbed metric.
The difference between these two is equal to the first order correction of
HEE $\Delta S_A$. Thus we do not need to know how the shape of the extremal surface is modified
under the metric perturbation. This is simply because of the fact that $\gamma_A$ satisfies the extremal surface
condition in background geometry.

Therefore we can calculate the shift of HEE $\Delta S_A$ due to the metric perturbation (\ref{pert})
from the formula  \be \Delta S_A=\f{1}{8G_N}\int
(d\zeta)^{d-1} \s{G^{(0)}}G^{(1)}_{\ap\beta}G^{(0)\ap\beta}, \label{dsa} \ee
where $\zeta$ is a coordinate of the $d-1$ dimensional extremal surface $\gamma_A$ as
employed in \cite{NNT}. $G^{(0)}$ and $G^{(1)}$ are defined by
\be
G^{(0)}_{\ap\beta}=\f{\de x^a}{\de \zeta^\ap}\f{\de x^b}{\de \zeta^\beta}g^{(0)}_{ab},\ \
\  G^{(1)}_{\ap\beta}=\f{\de x^a}{\de \zeta^\ap}\f{\de x^b}{\de \zeta^\beta}g^{(1)}_{ab},
\ee
and thus each of them represents the induced metric on $\gamma_A$ with respect to
the static background or its first order perturbation, respectively. Here $g^{(0)}$ and $g^{(1)}$ are
the metric of the background and that of its first order metric perturbation.

In this paper we concentrate on the examples where the subsystem $A$ is given by a round ball
for which we know the analytical expression of the surface $\gamma_A$ in the pure AdS. Note that
although we make specific symmetry assumptions about the background geometry \eqref{staticmet},
the fluctuations $h_{\mu \nu}$, that we shall consider here are completely arbitrary.
The main problem addressed in this paper is to ask what kind of differential equation
is satisfied by the shift of HEE $\Delta S_A$, assuming the normalizability near the AdS boundary $z\to 0$.

\section{HEE in BTZ black holes and Einstein Equation} \label{sec:BTZ}

Now we would like to focus on the lowest dimensional example: AdS$_3/$CFT$_2$.
Especially, we consider the pure gravity theory on a BTZ black hole, whose metric is given
by
 \begin{equation}\label{met0}
 ds^2 = \frac{R^2}{z^2} \left( -(1- m z^2) dt^2 + \frac{1}{(1-m z^2)} dz^2 + dx^2 \right).
 \end{equation}

We work in the GF-like gauge and consider metric perturbations of the following form
\begin{equation}\label{met1}
 ds^2_{(1)} = \frac{R^2}{z^2} \left(  h_{tt}(z,x,t)~ dt^2 + 2 h_{t x}(z,x,t)~ dt dx + h_{xx}(z,x,t)~ dx^2 \right)
\end{equation}
and we work up to linear order in $h_{\mu\nu}$.

\subsection{Linearized equations and their solution}
%
The linearized equations for $h_{xx}$, $h_{tx}$ and $h_{tt}$ are given by
\begin{equation}
 \begin{split}
  &\left( \partial_z - z~ \partial_z^2 \right) h_{tx} = 0, \\
  &\left( \partial_z - z~ (1 - m z^2)~ \partial_z^2 \right) h_{xx} = 0, \\
  &\left( 2 m^2 z^3 - (1-m z^2) \left( (1-2 m z^2) \partial_z - z (1- m z^2) \partial_z^2 \right)  \right) h_{tt} =0, \\
  & - 2 m z \partial_x h_{tx} + m z \partial_t h_{xx} - (1-mz^2) \left( \partial_z \left( \partial_x h_{tx}
  - \partial_t h_{xx} \right) \right)  = 0, \\
  & m z \partial_x h_{tt} + (1- m z^2 ) \left( \partial_z \left( \partial_x h_{tt} - \partial_t h_{tx} \right) \right) =0, \\
  &-2 m z h_{tt} + z \partial_x^2 h_{tt} - 2 z \partial_x \partial_t h_{tx} + z \partial_t^2 h_{xx}
  -(1- mz^2) \partial_z (h_{tt} - h_{xx}) =0 .
 \end{split}
\end{equation}
The solution of these set of equations are given by (keeping only the normalizable modes)
\begin{equation}\label{BTZlinsol}
\begin{split}
 h_{tx} &= z^2 \tilde H (x,t),\\
 h_{tt} &= \frac{2}{m} \left( \left(1-mz^2\right)^{\frac{1}{2}} - \left(1- m z^2\right) \right) H(x,t), \\
 h_{xx} &= \frac{2}{m} \left( 1 - \left(1-mz^2\right)^{\frac{1}{2}} \right) H(x,t).
 \end{split}
\end{equation}
where we must have
\begin{equation}
 \partial_t H = \partial_x \tilde H, ~~ \partial_t \tilde H = \partial_x H.
\end{equation}
and
\begin{equation}\label{Heq}
 \left( \partial_t^2 - \partial_x^2 \right) H = 0.
\end{equation}
%

\subsection{HEE due to metric perturbations}

We take the subsystem A to be the interval $\xi - \ell/2 \leq x \leq \xi + \ell/2$. Then the minimal surface
$\gamma_A$ is parameterized by a single coordinate $\phi$ as follows
\begin{equation}\label{surface}
 t = \text{constant}, ~ x = \xi + \frac{\ell}{2 \tanh ^{-1}{\beta}} \log \left[ \frac{\sqrt{1-\beta^2 \cos^2 \phi} - \beta \sin \phi}{\sqrt{1-\beta^2}}\right],
 ~ z =  \frac{\ell \beta }{2 \tanh ^{-1}{\beta}} \cos \phi.
\end{equation}
where $\beta$ is given in terms of $m$ as
\begin{equation}
 \beta = \tanh \left( {\frac{\ell \sqrt{m}}{2}} \right)
\end{equation}
and the coordinate $\phi$ takes values in $[-\pi/2, \pi/2]$.

Then the induced background metric \eqref{met0} and the induces first order metric \eqref{met1}
are given by
\begin{equation}\label{indmet}
 \begin{split}
  G^{(0)}_{\phi \phi} &= \frac{R^2}{\cos^2 \phi} \left( \frac{1}{1 - \beta^2 \cos^2 \phi} \right), \\
  G^{(1)}_{\phi \phi} &= R^2 \frac{h_{xx}}{1-\beta^2 \cos^2 \phi}
 \end{split}
\end{equation}
Now plugging \eqref{indmet} into \eqref{dsa} we have
\begin{equation} \label{EEbtzgen}
 \Delta S_A = \frac{R}{8 G_N} \int_{-\frac{\pi}{2}}^{\frac{\pi}{2}} d \phi
 \cos \phi \left( \frac{1}{\sqrt{1- \beta^2 \cos^2 \phi}} \right) h_{xx}.
\end{equation}
Plugging in
the solution of $h_{xx}$ from \eqref{BTZlinsol} into \eqref{EEbtzgen} we have
\begin{equation} \label{EEbtz}
 \Delta S_A(\xi, \ell, t) = \frac{R}{8 G_N} \int_{-\frac{\pi}{2}}^{\frac{\pi}{2}} d \phi ~F(\beta, \phi)
 ~H(x(\phi),t)
\end{equation}
where
\begin{equation}
\begin{split}
 F(\alpha, \phi) &= \frac{\ell^2}{4 (\tanh^{-1} \beta)^2} \left( \frac{1- \sqrt{1-\beta^2 \cos^2 \phi}}{\sqrt{1- \beta^2 \cos^2 \phi}} \right)\cos \phi
 \\
\end{split}
\end{equation}
and in \eqref{EEbtz} $x$ in the argument of the function $H$ is to be substituted in terms of $\phi$ using \eqref{surface}.

Putting these together,
\begin{equation}
\Delta S_A(t,\xi,l)=\frac{R}{4m G_N}\int^{\frac{\pi}{2}}_{-\frac{\pi}{2}}d\phi
\cos\phi\left(\frac{1}{\sqrt{1-\beta^2\cos^2\phi}}-1\right)H(t,x),
\end{equation}

where note that
\begin{equation}\label{}
\begin{split}
& \beta=\tanh\frac{l\sqrt{m}}{2},\\
& x=\xi+\frac{1}{\sqrt{m}}\log \left(\frac{\sqrt{1-\beta^2\cos^2\phi}-\beta\sin\phi}{\sqrt{1-\beta^2}}\right).
\end{split}
\end{equation}

After the Fourier transformation w.r.t $\xi$ and perform the integral of $\phi$ explicitly, we get
\begin{equation}\label{EEfinexp}
 \begin{split}
 \Delta S_A(t,k,l) &=\frac{R}{4mG_N}H(t,k)\cdot \int^{\frac{\pi}{2}}_{-\frac{\pi}{2}}d\phi
\cos\phi\left(\frac{1}{\sqrt{1-\beta^2\cos^2\phi}}-1\right)e^{i\frac{k}{\sqrt{m}}\log \left(\frac{\sqrt{1-\beta^2\cos^2\phi}-\beta\sin\phi}{\sqrt{1-\beta^2}}\right)},\\
& =\frac{R}{2G_N}\cdot\frac{H(t,k)}{k^2+m}\cdot\left(-\cos \left(\frac{kl}{2}\right)+\frac{\sqrt{m}}{k\beta}\sin \left(\frac{kl}{2}\right)\right).
\end{split}
\end{equation}

\subsection{Equations satisfied by HEE}

Acting on \eqref{EEbtz} by the operator $(\partial_t^2 - \partial_\xi^2)$ and using \eqref{Heq} we can easily conclude
\begin{equation}\label{Heeq}
 \left( \partial_t^2 - \partial_\xi^2 \right) \Delta S_A (t, \xi, l) = 0.
\end{equation}
Also we can easily see from \eqref{EEfinexp} that $\Delta S_A(t,\xi,l)$ satisfies
\begin{equation}\label{Heeeq}
\left(\partial_l^2-\frac{1}{4}\partial_\xi^2-\frac{m}{2\sinh^2(l\sqrt{m}/2)}\right)\Delta S_A(t,\xi,l)=0.
\end{equation}

Indeed, these relations agree with the known results \cite{NNPT} in the pure
AdS$_3$ limit $m\to 0$. The result of \cite{NNPT} for the pure AdS$_3$ was later
reproduced purely from CFT calculations in \cite{Wong:2013gua}. Thus we expect that our
results (\ref{Heeq}) and (\ref{Heeeq}) can be confirmed in a similar way.

If we take the large size limit $l\to \infty$, we find that the potential term in (\ref{Heeeq}) gets vanishing. If we assume the translational invariance, then we immediately obtain $\Delta S_A\propto l$. This agrees with the extensive part of entanglement entropy, which typically appears for mixed states.

In this way we derived an analogue of Einstein equation (\ref{Heeq}) and (\ref{Heeeq}) for a pure gravity on the BTZ black hole in terms of entanglement entropy. One may notice (\ref{Heeq}) determines the time evolution of $\Delta S_A$. However, this is a special feature in three dimensional gravity, where there are no propagating gravitational waves. Indeed, as we will later, in higher dimensional gravity, we will only obtain a counter part of (\ref{Heeeq}).

\section{Analysis in AdS4}\label{sec:AdS4}

\subsection{Calculations of HEE}

Now we would like to move on to a higher dimensional example. We analyze the shift of
HEE $\Delta S_A$ in the presence of metric perturbation $h_{\mu\nu}$ described by (\ref{adsmet})
and (\ref{pert}), especially in the pure gravity theory on AdS$_4$.
Though we focus on this four dimensional example, we can generalize our results shown below to higher
dimensional examples in a straightforward way. We consider the modes which are normalizable near the
AdS boundary $z\to 0$, where the metric perturbation looks like in the Fourier basis:
\be
h_{\mu\nu}(t,x,y,l)=e^{-\omega t+ik_x x+ik_y y}h_{\mu\nu}(k_x,k_y,\omega)(\omega^2-k^2)^{-3/4}z^{3/2}J_{3/2}(\s{\omega^2-k^2}z).
\ee
We also define $H_{\mu\nu}$ which is obtained from the near boundary expansion of $h_{\mu\nu}$ in the $z\to 0$ limit:
\be
h_{\mu\nu}(t,x,y,z)=z^3\cdot H_{\mu\nu}(t,x,y)+O(z^4).
\ee
The holographic energy stress tensor \cite{EMT} (i.e. the energy stress tensor in the dual CFT) is given by
\be
T_{\mu\nu}=\f{3R^{d-1}}{16\pi G_N}H_{\mu\nu}.
\ee

We consider the shift of HEE $\Delta S_A$ by choosing
 the subsystem $A$ to be a disk with a radius $l$. In the pure AdS$_4$, the HEE is computed as the area of the minimal surface $\gamma_A$ given by the half of sphere parameterized by
\ba
x=l\sin\theta \cos\phi+X,\ \ \  y=l\sin\theta \sin\phi+Y, \ \ \ z=l\cos\theta,
\ea
with the range $0<\theta<\pi/2$ and $0<\phi<2\pi$.

$\Delta S_A$ is calculated by using (\ref{dsa}) as follows:
\ba
&& \Delta S_A(t,X,Y,l)\no
&&=\f{R^2}{8G_N}\!\int^{2\pi}_0\!\! d\phi\!\int^{\pi/2}_0\!\! d\theta\!
\f{\sin\theta}{\cos^2\theta} \left[(1\!-\!\sin^2\theta \cos^2\phi)h_{xx}-2h_{xy}\sin^2\theta\cos\phi\sin\phi
+(1\!-\!\sin^2\theta\sin^2\phi)h_{yy}\right].\label{HEFS}  \no
\ea

It is useful to take the Fourier transformation of $\Delta S_A(t,X,Y,l)$ with respect to
$t,X,Y$, which is denoted by $\Delta S_A(\omega,k_x,k_y,l)$. Then we find (see \cite{NNPT} for more details)
\ba
&& \Delta S_A(\omega,k_x,k_y,l) \no
&& =\f{3R^2}{8G_N}\s{\f{\pi}{2}}\int^{\pi/2}_0 d\theta \f{\sin\theta}{\cos^2\theta}
\int^{2\pi}_0 d\phi I(\phi,\theta)\cdot e^{il\sin\theta(k_x\cos\phi+k_y\sin\phi)}\no
&&\ \ \ \ \ \times \left(\f{l\cos\theta}{\s{\omega^2-k^2}}\right)^{3/2}\cdot J_{3/2}(l\cos\theta\s{\omega^2-k^2}).\ \ \  \label{HEEFR}
\ea
We defined
\ba
I(\phi,\theta)&=&(1-\sin^2\theta \cos^2\phi) H_{xx}(\omega,k_x,k_y)-2\sin^2\theta\cos\phi\sin\phi H_{xy}(\omega,k_x,k_y) \no
&&+(1-\sin^2\theta\sin^2\phi) H_{yy}(\omega,k_x,k_y),
\ea
where $H_{\mu\nu}(\omega,k_x,k_y)$ denote the Fourier transformation of $H_{\mu\nu}(t,x,y)$.

By performing the integrals and employing the Einstein equation, we can express $\Delta S_A(\omega,k_x,k_y,l)$ in terms of the (holographic) energy
stress tensor $T_{tt}(\omega,k_x,k_y)$ as follows \cite{NNPT}:
\ba
\Delta S_A(\omega,k_x,k_y,l)=\f{2\s{2}\pi^{5/2}l^{3/2}}{(\omega^2-k^2)^{3/4}}\cdot T_{tt}(\omega,k_x,k_y)\cdot \int^{\pi/2}_0 d\theta \f{\sin\theta}{\s{\cos\theta}}Q(\theta), \label{Trel}
\ea
where $Q(l)$ is defined by
\be
Q(\theta)=J_{3/2}(l\cos\theta\s{\omega^2-k^2})\cdot \left[\left(1-\f{\sin^2\theta}{2}\right)J_0(kl\sin\theta)
+\f{\sin^2\theta}{k^2}\left(\omega^2-\f{k^2}{2}\right)J_2(kl\sin\theta)\right].
\ee
As shown in \cite{Blanco:2013joa}, we can perform the integral of $\theta$ in
(\ref{Trel}) analytically and finally we obtain
\be
\Delta S_A(k_x,k_y,\omega,l) = 4 \pi^2 \f{l}{k^2}J_2(kl)T_{tt}(k_x,k_y,\omega).
\label{fHEE}
\ee
This result again agrees with the result from CFT calculations as confirmed in \cite{Blanco:2013joa}.

\subsection{Equation Satisfied by HEE}

Thus we can show that the perturbative Einstein equation requires that $\Delta S_A$ (\ref{fHEE})
in pure gravity on the AdS$_4$ satisfies the following differential equation, assuming the normalizability
condition near the AdS boundary:
\be
\left[\f{\de^2}{\de l^2}-\f{1}{l}\f{\de}{\de l}-\f{3}{l^2}-\f{\de^2}{\de x^2}-\f{\de^2}{\de y^2}\right]\Delta S_A(t,x,y,l)=0.  \label{xx}
\ee
 Remember that $\Delta S_A$ describes how much the entanglement entropy increases when we excite a ground state of a given CFT.

At first one may expect that the Einstein equation leads to something like a differential equation which looks like a scalar field on AdS$_{4}$. However, this naive guess is not
completely correct. We notice that this equation (\ref{xx}) does not involve any time derivatives and looks like a constraint equation. Instead, the size $l$ of the subsystem plays the role of time and our differential equation is hyperbolic. Therefore it is much like a scalar field equation on AdS$_3$. We will discuss the interpretation of this equation soon later.

When the excitations are translationally invariant\footnote{For more generic choices of the subsystem $A$ such as the strips, we need to also require the rotational invariance to  obtain the first law-like relation as pointed out in \cite{Allahbakhshi:2013rda,Guo:2013aca,Blanco:2013joa}. See also \cite{NNT,Caputa:2013eka} for other developments on the first law-like relation.}, we know that $\Delta S_A$  satisfies the first law-like relation
\be
\Delta S_A=\f{\Delta E_A}{T_{ent}}=\f{\pi^2 l^3}{2}T_{tt},
\ee
where $E_A=\pi l^2T_{tt}$ is the total energy included in $A$ and $T_{ent}$ is given by
$T_{ent}=\f{2}{\pi l}$ as found in \cite{BNTU}. Our constraint equation (\ref{xx}) is consistent with this fact.

It is also worth mentioning how the equation (\ref{xx}) is modified when we consider
an Einstein gravity coupled to some matter fields. Assuming the matter contributions
as small as the metric perturbation, this problem was discussed in \cite{NNPT} for the
AdS$_3$ setup. It is straightforward to extend this result to our higher dimensional example.
In the end we find that in the presence of matter fields in gravity, (\ref{xx}) is modified as follows
\be
\left[\f{\de^2}{\de l^2}-\f{1}{l}\f{\de}{\de l}-\f{3}{l^2}-\f{\de^2 }{\de x^2}-\f{\de^2 }{\de y^2}\right]\Delta S_A(t,x,y,l)=\la O\lb \la O\lb  .  \label{xxy}
\ee
$\la O\lb$ means expectation values of operators dual to matter fields and
the term $\la O\lb \la O\lb$ describes abstractly a rather complicated sum of bilinear
terms of one point functions with various coefficients.

\subsection{IR Boundary Condition}

Since (\ref{xx}) is independent from the time derivatives of $\Delta S_A$, one may wonder how we can determine its time evolution.
This problem is resolved if we remember that we need to impose an appropriate IR boundary condition
in addition to the UV boundary condition (i.e. normalizability) which we already employed. This is because the
IR boundary condition leads to a dispersion relation $\omega=\omega(k_x,k_y)$.

A simple example will be a hard wall cut off in the IR region at $z=z_0$. We impose the
metric perturbation is vanishing at $z=z_0$.  This requires
\be
J_{3/2}(z_0\s{\omega^2-k^2})=0.
\ee
This can be solved as
\be
k^2=\omega^2-\omega_n^2, \ \ \ (n=1,2,\ddd).
\ee
For each mode, we find that the EOM (\ref{xx}) is rewritten as follows:
\be
\left[\f{\de^2}{\de l^2}-\f{1}{l}\f{\de}{\de l}-\f{3}{l^2}-\f{\de^2}{\de t^2}-\omega_n^2\right]\Delta S_A=0.  \label{xy}
\ee
This equation looks more intuitive as it is a hyperbolic differential equation where $l$ and $t$ played role of space and time, respectively.

Another typical example of IR boundary condition in AdS/CFT is the presence of a horizon, where we impose the ingoing boundary condition so that the setup follows a retarded propagation. In this case we can obtain the dispersion relation as that of the quasi normal mode, which constraints the HEE as follows:
\be
\left[i\f{\de}{\de t}-\omega\left(-i\f{\de}{\de x},-i\f{\de}{\de y}\right)\right]
\Delta S_A(t,x,y,l)=0.
\ee
Since the dispersion relation is clearly independent from $l$, this time evolution commutes with the constraint (\ref{xx}). This completely determines the time evolution of HEE. We will study an explicit example in the next section.

\subsection{An Interpretation of Constraint Equation}

 In summary, we found that the perturbative Einstein equation plus the normalizability condition near the AdS boundary is equivalent to the constraint equation
(\ref{xx}) in the AdS$_4$ pure gravity setup. The normalizability simply means that we are considering an excited state in the dual CFT without changing the original lagrangian.
This result (\ref{xx}) or (\ref{xxy}) can be applied to the near boundary region of any asymptotically AdS spacetime with an arbitrary IR geometry.

\begin{figure}[ttt]
   \begin{center}
     \includegraphics[height=4cm]{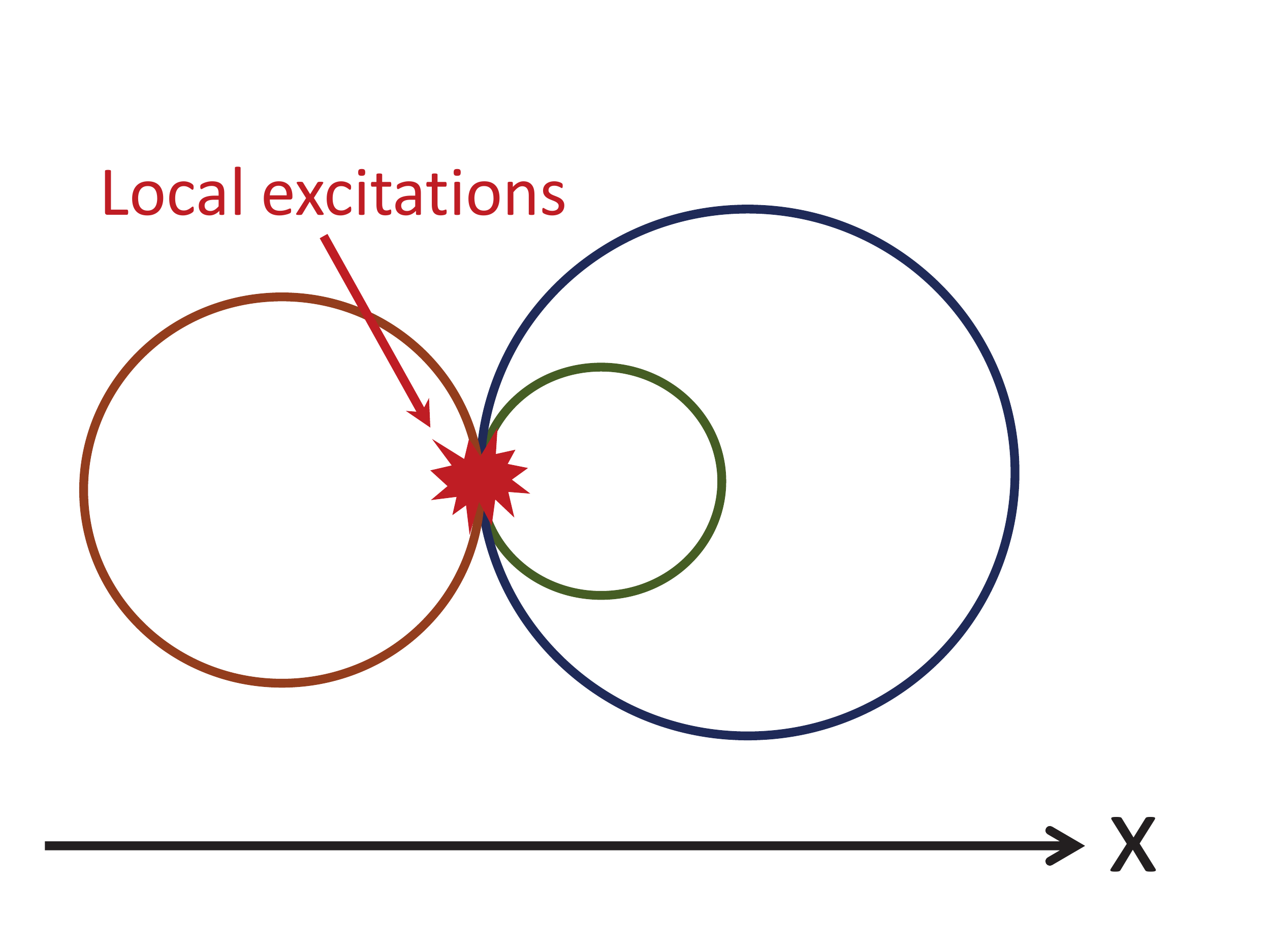}
   \end{center}
   \caption{The schematic picture of the interpretation of constraint equation.
    The HEE becomes non-trivial only if the boundary $\de A$ of the subsystem $A$ intersects with locally excited regions. The brown, green and blue circles are three examples of the boundaries $\de A$ for which the HEE becomes non-trivial.}\label{fig:loc}
\end{figure}

As already mentioned, it may sound surprising that the equation (\ref{xx}) or (\ref{xxy}) does not determine the time evolution but offers a constraint equation of $\Delta S_A$
at a fixed time. Here we would like to propose an interpretation that it gives an consistency condition which should be satisfied by the entanglement entropy. Even though we cannot explain precisely the form of (\ref{xx}), we can understand why it takes the form of a hyperbolic differential equation. If we roughly regard (\ref{xx}) as a wave equation, then $l$ corresponds to a `time' and $r=\s{x^2+y^2}$ does to the radial coordinate. The wave equation tells us that a perturbation propagates at the `speed of light'. So if we perturb $r=0$ at $l=0$, then the perturbation is localized at the region $l=r$. In fact, this fact is naturally expected for the entanglement entropy. When we excite only a small region, the entanglement entropy $S_A$ will be increased only if the boundary of the subsystem $A$ crosses this small region, which is essentially the same as the statement $l=r$ (see Fig.\ref{fig:loc}). This argument suggests that the values of $\Delta S_A(t,x,y,l)$ for different values of $(x,y,l)$ at a fixed
time $t$ should be related in the manner explained. In this way, we can understand (\ref{xx}) as this consistency condition. A more general equation (\ref{xxy}) described that perturbations of matter fields also contribute to $\Delta S_A$.

\section{HEE for AdS4 BH in the Large Subsystem Size Limit}\label{sec:AdS4BH}

In this section we shall consider the $AdS_4$ Schwarzschild black hole and long-wavelength fluctuations about it. 
To simplify our analysis we will also assume the size of the subsystem to be very large compared to length 
scale associated to the temperature of the system.

\subsection{Metric Perturbation in AdS BH}

The metric of AdS$_4$ black hole is given by
\be
ds^2_{AdSBH}=-\f{f(z)}{z^2}dt^2+\f{1}{z^2f(z)}dz^2+\f{1}{z^2}(dx^2+dy^2), \label{bhmt}
\ee
where $f(z)=1-(z/z_H)^3$. We have set $R=1$ just for simplicity. We choose the subsystem $A$ to be a round disk with the radius $l$. Its center is specified by the values of $(t,x,y)$.

We are interested in the metric perturbation around this background, which can be described by
\be
ds^2=ds^2_{AdSBH}+\f{h_{\mu\nu}dx^\mu dx^\nu}{z^2}, \label{pertbh}
\ee
where $\mu=t,x$ and $y$.

In this section we would like to analyze the behavior of HEE under this metric perturbation and obtain the constraint equation imposed by the pure gravity Einstein equation. If we take the small size limit $l\to 0$, the problem is reduced to the previous result (\ref{xx}) for the pure AdS$_4$. Therefore, in order to find analytical results, we will take the large size limit $l\to \infty$, where the black hole effect becomes maximally important.

\subsection{Minimal Surface Without Metric Perturbations}

As a preparation for the calculation of $\Delta S_A$, we would like to work out the profile of minimal surfaces in the (unperturbed) AdS$_4$ black hole. Consider the large size limit $l\to \infty$, for a spherical subsystem $A$ with radius $l$. In this limit, the minimal surface extends from the AdS boundary $z=0$ to the near horizon region $z\simeq z_H$ and almost wraps the horizon. We write the maximum of $z$ coordinate on this minimal surface as $z=z_*$, where $z_*$ approaches $z_H$ in the limit $l\to \infty$. We explained the details of minimal surface calculations in the appendix A.

\begin{figure}[ttt]
   \begin{center}
     \includegraphics[height=4cm]{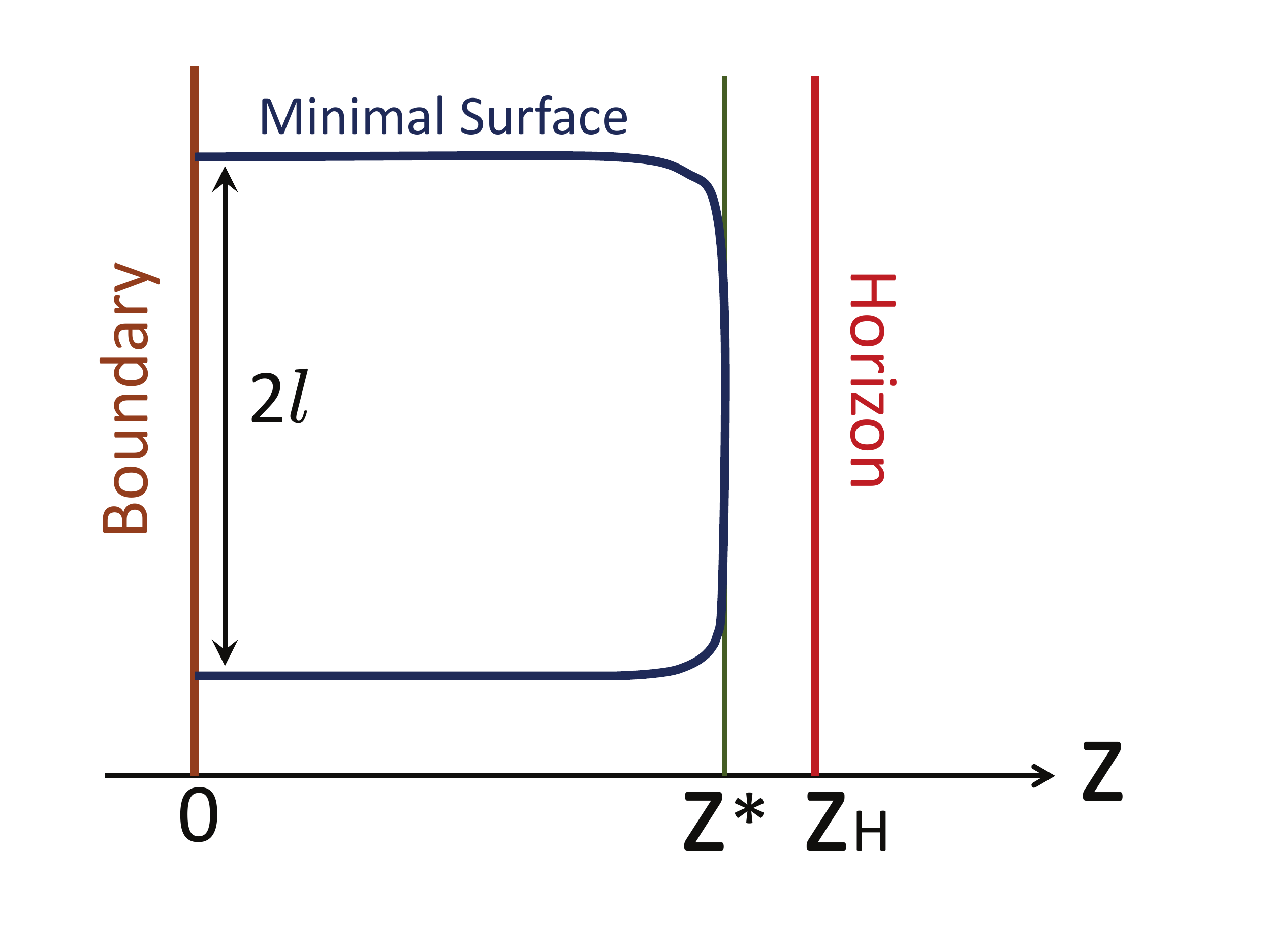}
   \end{center}
   \caption{A sketch of the minimal surface in the AdS BH when the size $l$ of the subsystem is very large.}\label{fig:min}
\end{figure}

Thus we can approximate the corresponding minimal surface by a cylinder
with its bottom on one side (refer to Fig.\ref{fig:min} for its sketch).
In other words, this is the union of two dimensional manifolds $P_A$
and $Q_A$. $P_A$ is the cylinder defined by
\be
(t,x,y,z)=(t_0,l\cos\theta,l\sin\theta,\xi),
\ee
where $0\leq \xi\leq z_*$ and $0\leq \theta\leq 2\pi$.  The other one $Q_A$ is the round disk given by
\be
0\leq x^2+y^2\leq l^2, \ \ t=t_0,\ \ \ z=z_*.
\ee

We can numerically calculate the maximal value $z_*$ of $z$ on the minimal surface as a function of $l$ in the appendix A. We can confirm the following relation in $l\to \infty$ limit for AdS4:
\be
z_*(l)\simeq z_H-A\cdot e^{-\beta l/z_H}, \label{rela}
\ee
where $A$ and $\beta$ are constants
\footnote{In the AdS3 black hole (BTZ black hole), we can analytically show the relation $z_*\simeq z_H-2z_He^{-2l/z_H}$, 
i.e. $\beta=2$, where the length of the interval $A$ is defined to be $2l$ instead of $l$. Furthermore, our numerical analysis 
confirms $\beta=4$ for AdS$_5$. This suggest that for AdS$_{d+1}$ we get $\beta=d$. For the 
case of a strip subsystem, a similar expression was obtained in \cite{Fischler:2012ca}.}. Moreover, our numerical results show
\be
\beta=3. \label{bett}
\ee

\subsection{HEE with the metric perturbations}

Now we would like to calculate $\Delta S_A$ in the presence of the metric perturbations (\ref{pertbh}). In the large $l$ limit, there are two contributions
$\Delta S_A^P$ and $\Delta S_A^Q$ from the two parts of the minimal surface $P_A$ and
$Q_A$:
\be
\Delta S_A=\Delta S^P_A+\Delta S^Q_A. \label{sabv}
\ee

They are computed as follows:
\ba
\Delta S_A^{P}&=&\f{1}{8G_N}\int^{z_*}_0\f{l}{z^2\s{f(z)}}\int^{2\pi}_0 d\theta\left[
h_{xx}\sin^2\theta-2h_{xy}\sin\theta\cos\theta+h_{yy}\cos^2\theta\right]
e^{il(k_x\cos\theta+k_y\sin\theta)}  \no
&=&\f{2\pi l}{8G_N}\int^{z_*}_0\! \f{dz}{z^2\s{f(z)}}\!\left[\f{h_{xx}+h_{yy}}{2}J_0(kl)+
\f{k_x^2-k_y^2}{2k^2}(h_{xx}-h_{yy})J_{2}(kl)+\f{2k_xk_y}{k^2}h_{xy}J_{2}(kl)\right],\no
&& \label{delp} \no
\Delta S_A^{Q}&=&\f{1}{8G_Nz_*^2}\int^{l}_0 rdr \int^{2\pi}_0 d\theta e^{ir(k_x\cos\theta+k_y\sin\theta)}
[h_{xx}+h_{yy}]_{z=z_*}\no
&=& \f{1}{8G_N}\cdot \f{lJ_1(kl)}{z_*^2 k}\cdot [h_{xx}+h_{yy}]_{z=z_*}.
\label{delq}
\ea
We already took the Fourier transformation of $h_{\mu\nu}$ w.r.t $(x,y,t)$ and their momenta are described by
$k_x,k_y,\omega$.  Thus the metric perturbation $h_{\mu\nu}$ in the above only depends on $z$.

\subsection{Metric perturbations}

Using the U(1) rotation symmetry of the background geometry we can set $k_y=0$, while
 $k_x\neq 0$. Then using the perturbative solution to Einstein equation in the 
 long-wavelength limit with the vanishing
 energy stress tensor, in the sound mode \cite{Kovtun:2005ev}, we get
\ba
&& h_{xx}+h_{yy}=2C_{y}\s{1-z^3/z_H^3}+C_{x},\no
&& h_{xx}-h_{yy}=C_x-C_{+} (1-z^3/z_H^3)^{i\f{z_H\omega}{3}}-C_{-}(1-z^3/z_H^3)^{-i\f{z_H\omega}{3}},  \label{pmw}
\ea
with non-trivial values of $h_{tt}$ and $h_{xt}$. The constants $C_{x},C_y,C_+$ and
$C_{-}$ are arbitrary constants. If we require that the mode is normalizable near the AdS boundary, we find
\be
C_x=-2C_y,\ \ \  C_x=C_{-}+C_{+}. \label{nomw}
\ee

We will not consider the sheer mode here because this does not contribute to $\Delta S_A$.

\subsection{Analysis of HEE}

Now we would like to evaluate (\ref{sabv}) in the presence of metric perturbations.
We require that the metric perturbation is normalizable as in (\ref{nomw}) because we need to keep the boundary metric flat.
First of all, it is immediate to see from (\ref{delp}) that the shear mode does not affect $\Delta S_A$ by setting $k_y=0$ using the rotational symmetry. Therefore we can indeed focus on the sound mode which is given by (\ref{pmw}).
In this way we find
\ba
\Delta S^P_A&=&\f{\pi l}{8G_N}\int^{z_*}_0 \f{dz}{z^2\s{f(z)}}\left(H_{+}(z)J_0(kl)+
H_{-}(z)J_{2}(kl)\right),\no
\Delta S_A^{Q}
&=& \f{\pi l}{4G_N}\cdot \f{J_1(kl)}{z_*^2 k}\cdot H_{+}(z_*),  \label{ppp}
\ea
where
\ba
&& H_{+}(z)=h_{xx}+h_{yy}=(C_+ + C_{-})\left(1-\s{1-z^3/z_H^3}\right),\no
&& H_{-}(z)=h_{xx}-h_{yy}=C_+\left(1-(1-z^3/z_H^3)^{i\f{z_H\omega}{3}}\right)
+C_-\left(1-(1-z^3/z_H^3)^{-i\f{z_H\omega}{3}}\right).\no
\ea

The large $l$ limit means $l>>z_H$ and note also that we assumed the hydrodynamical limit $kz_H<<1$ and $\omega z_H<<1$ to find (\ref{pmw}). Remembering $z_*\simeq z_H$ in the large $l$ limit as in (\ref{rela}),
we can easily estimate the order of (\ref{ppp}) as follows
\be
\Delta S_A^P\sim \f{l}{G_Nz_H}f_P(kl),\ \ \
 \Delta S_A^Q\sim \f{l}{G_N k z_H^2}f_Q(kl).
\ee
Thus in the hydrodynamics limit, it is clear
\be
\Delta S_A^P<<\Delta S_A^Q.
\ee

In this way we finally obtain the leading contribution:
\be
\Delta S_A \simeq \Delta S_A^Q=\f{\pi lJ_1(kl)}{4G_Nk z_H^2}(C_++C_{-}). \label{heef}
\ee

Therefore we can conclude that in the $l\to \infty$ limit, $\Delta S_A$ satisfies
the following constraint equation
\be
\left[\f{\de^2}{\de l^2}-\f{1}{l}\f{\de}{\de l}-\f{\de^2 }{\de x^2}-\f{\de^2 }{\de y^2}\right]\Delta S_A=0. \label{adsbhem}
\ee

Interestingly, this coincides with the analogous result for the flat space holography (\ref{flatee}), where we computed the HEE $\Delta S_A$ from a perturbation on a flat spacetime as we discuss in the appendix B. This fact may be natural because the near horizon limit of the AdS BH is given by the Rindler geometry, which is equivalent to the flat Minkowski spacetime. Both follow the volume law of entanglement entropy instead of area law. Indeed, if we assume the translational invariance, immediately we obtain
$\Delta S_A\propto l^2$ from (\ref{adsbhem}), which is proportional to volume. On the other hand, the holographic calculation clearly shows the volume law for the Einstein  gravity on a flat space \cite{LiTa,NRT}.

\subsection{Comments on IR Boundary Condition}

For a causal time evolution, we require the ingoing boundary condition at the horizon. For the metric field, this is equivalent to setting $C_+=0$. Here we would like consider how this IR boundary condition is interpreted as the boundary condition of $\Delta S_A$.

This boundary condition does not affect the leading behavior of $\Delta S_A$, which satisfies the constraint (\ref{adsbhem}). However, this ingoing condition is non-trivial if we consider the subleading time-dependent term of $\Delta S_A$ in the $l\to\infty$ limit. We can see from $\Delta S_A^P$ in (\ref{ppp}) that sub-leading time-dependent contribution of $\Delta S_A$ behaves like
\ba
&& \mbox{Ingoing}\ \  (C_+=0): \ e^{-i\omega t}(z_*-z)^{1- \f{i\omega z_H}{3}}\propto e^{-3l/z_H}\cdot e^{-i\omega (t- l)}, \\
&& \mbox{Outgoing}\ \  (C_-=0): \ e^{-i\omega t}(z_*-z)^{1+ \f{i\omega z_H}{3}}\propto e^{-3l/z_H}\cdot e^{-i\omega (t+l)},
\ea
where we used (\ref{bett}). After the integration over $\omega$ with some Fourier weight, we find that for the ingoing boundary condition, the term which looks like $e^{-3l/z_H}\cdot F(t-l)$ is allowed, while the one like $e^{-3l/z_H}\cdot F(t+l)$ is excluded. In this way, we managed to find the entropic counterpart of the IR boundary condition for the AdS black hole.

\section{Boosted Subsystem} \label{sec:boost}

So far we discussed an analogue of Einstein equation for the HEE $\Delta S_A$. However, only one linear combination of components of the metric perturbation contributes to $\Delta S_A$.
To pick up all independent components of metric perturbations, we need to consider some other quantities. In this section, we would like to point out that we can employ the HEE for a boosted subsystems for this purpose.

Let us focus on the example of the pure Einstein gravity on AdS$_4$.
Before we boost the subsystem $A$, $\Delta S_A$ is given by (\ref{fHEE}).
Now we boost the subsystem $A$. The corresponding minimal surface is given by
\ba
&& t=\sinh\beta l \sin\theta \cos\phi +t_0,\no
&& x=\cosh\beta l \sin\theta \cos\phi +x_0,\no
&& y=l \sin\theta \sin\phi +y_0,\no
&& z=l \cos\theta.
\ea

If we define the unboosted coordinate by
\be
(\ti{t},\ti{x})=(\cosh\beta t-\sinh\beta x,-\sinh\beta t+\cosh\beta x), \label{boost}
\ee
we obtain
\be
\Delta S_A\propto \f{l}{\ti{k}^2}J_2(\ti{k}l)T_{\ti{t}\ti{t}}(\ti{k},\ti{\omega}),
\ee
where
\be
T_{\ti{t}\ti{t}}=\sinh^2\beta T_{xx}-2\sinh\beta\cosh\beta T_{tx}+\cosh^2\beta T_{tt}.
\ee

In this way, $\Delta S^{(\beta)}_A$, which is the HEE for a boosted subsystem with the parameter $\beta$, satisfies the following the differential equation:
\be
\left[\f{\de^2}{\de l^2}-\f{1}{l}\f{\de}{\de l}-\left(\cosh\beta \f{\de}{\de x}+\sinh\beta \f{\de}{\de t}\right)^2-\f{\de^2 }{\de y^2}-\f{3}{l^2}\right]\Delta S^{(\beta)}_A(t,x,y,l)=0,
\ee
where notice the relation
\be
\f{\de}{\de \ti{x}}=\cosh\beta \f{\de}{\de x}+\sinh\beta \f{\de}{\de t}.
\ee

Note that only two out of six components of $T_{ab}$ are independent due to the equation of motion. Therefore in this pure gravity model, there are only two independent quantities, which can be chosen to be $\Delta S^{(\beta_1)}_A$ and $\Delta S^{(\beta_2)}_A$ for two different values $\beta_1$ and $\beta_2$.

\section{Conclusions}

 The main purpose of this paper is to find a gauge invariant reformulation of Einstein equation in the context of AdS/CFT.
 Even though the metric itself is not gauge invariant, the HEE is an gauge invariant quantity in gravitational theories.
 Note that here the existence of the boundary of AdS spacetime is crucial in that it provides us with the fixed flat
 coordinate system $(t,x,y)$. In the full AdS geometry, still we have a degrees of freedom to modify the extra dimension
 coordinate $z$. However, this ambiguity is removed once we focus on the HEE $\Delta S_A$ because we can use the parameter $l$,
 which is the size of the subsystem, instead of $z$. Interestingly, the HEE is gauge invariant also from the CFT side and is a
 nice physical quantity in the dual CFT.

In this paper we first work out the entropic counterpart of Einstein equation for the three dimensional gravity
on BTZ black holes. This is given by two differential equations. One of them (\ref{Heeq}) is the wave equation of the
real time $t$ and the center position $\xi$ of the subsystem $A$. Another one (\ref{Heeeq}) is a hyperbolic differential
equation about the size $l$ and the position $\xi$, which looks like a constraint equation at each time because the
time derivatives are not included in this equation. The difference between the pure AdS$_3$ and the BTZ black hole
reflects in the potential term of the hyperbolic differential equation.

In this three dimensional example, the equations constraint the time evolution of $\Delta S_A$ owing to the first equation.
However, this is not true in higher dimensional example (especially the AdS$_4$ example) and there is no counter part
of this first equation. Even though the Einstein equation describes the time evolution of the metric, it leads to a constraint
type equation (\ref{xx}) at each time for $\Delta S_A$ without time derivatives. In this sense we can say that the three dimensional
result is special in that there is no propagating gravitational wave.

We gave a heuristic interpretation of the constraint equation we got from the Einstein equation as a consistency
equation of entanglement entropy for various sizes and positions. The time evolution of $\Delta S_A$ is fixed by the
IR boundary condition in AdS/CFT, which leads to a dispersion relation. Thus there is no contradiction with what we normally
expect in AdS/CFT.

We also examined how our result is modified for AdS$_4$ black holes.  The effect of presence of horizon is particularly
emphasized in the $l\to \infty$ limit and we studied this limit. In the end, we find that analogue of Einstein equation is
now written as (\ref{adsbhem}). In addition we studied a subleading contribution in the  
$l\to \infty$ limit and found an interesting connection to quasi normal modes and found that this gives IR boundary condition for the HEE dynamics. It might be interesting to find any relation between this and the scaling property found in \cite{Liu:2013iza} where the time-evolution of entanglement entropy was studied in detail for a setup of quantum quenches. 

Moreover, we noted that $\Delta S_A$ itself is not enough to equivalently describe all
degrees of freedom of metric perturbation, even if we impose the on-shell condition.
To resolve this problem we introduced the entanglement entropies for boosted subsystems and showed that they offer a complete basis to cover all metric components in AdS$_4/$CFT$_3$ setup.

It will be an important future problem how to generalize our result to full non-linear Einstein equation.
It may be also intriguing to consider an action formalism of Einstein gravity using $\Delta S_A$ as a fundamental
field with the IR boundary condition taken into account appropriately.

\acknowledgments
{
We would like to thank Thomas Faulkner, Gary Horowitz, Veronika Hubeny, Juan Maldacena,
Robert Myers, Mukund Rangamani and Shinsei Ryu for useful discussions. TT is very grateful to the organizers
of strings 2013 conference held in Sogang University, Seoul and organizers of GR20 conference held
in Uniwersytet Warszawski, Warsaw for their hospitalities, where a part of this work was presented.
We would also like to thank the organizers of Gauge/Gravity Duality 2013 workshop held at Max Planck
Institute for Physics in Munich, where this work was presented and was almost completed. TT is supported
by JSPS Grant-in-Aid for Challenging Exploratory Research No.24654057 and JSPS Grant-in-Aid for Scientific
Research (B) No.25287058. We are also supported by World Premier International
Research Center Initiative (WPI Initiative) from the Japan Ministry
of Education, Culture, Sports, Science and Technology (MEXT).
}

\appendix

\section{Analysis of Minimal Surface in AdS4 BH}
Here we analyze the minimal surfaces in the AdS$_4$ black hole (\ref{bhmt}) in the large
subsystem size limit $l\to\infty$.

We are interested in minimal surfaces of the form
\be
x^2+y^2=r(z)^2,\ \ \ t=t_0.
\ee
We need to minimize the area functional
\be
\mbox{Area}=\int dz \f{r(z)}{z^2}\s{(\de_z r(z))^2+\f{1}{f(z)}}.
\ee
The equation of minimal surface is given by
\be
\f{1}{z^2}\s{(\de_z r(z))^2+\f{1}{f(z)}}=\de_z\left[\f{r(z)\de_z r(z)}{z^2\s{(\de_z r(z))^2+\f{1}{f(z)}}}\right].
\ee

We plotted several examples of the function $r=r(z)$ as a function $z$ in Fig.\ref{fig:BHE} below.
As $l$ gets larger  or equally as $z_*$ gets closed to $z_H$ (see Fig.\ref{fig:Zl}), the minimal surface
approaches to the cylinder because $r(z)$ becomes a constant function.

\begin{figure}[ht]
\begin{minipage}[b]{0.45\linewidth}
\centering
\includegraphics[width=\textwidth]{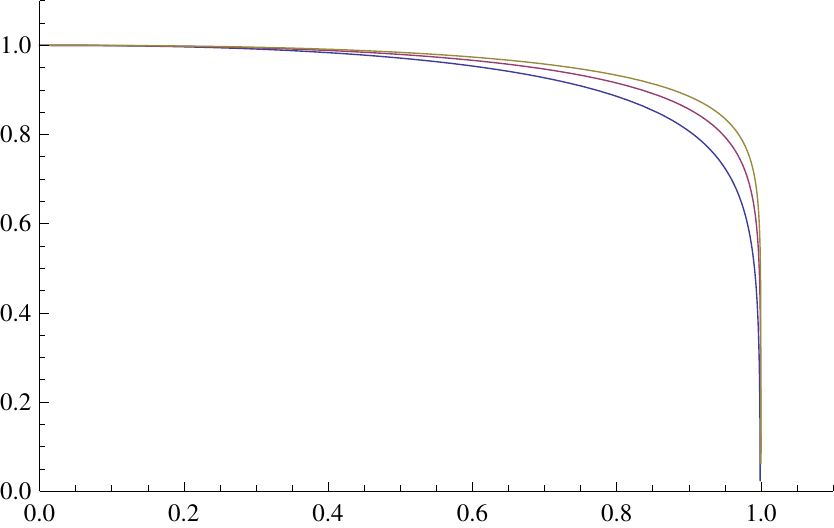}
\caption{The plot of $r(z)/l$ for various choices of $z_*$ (or equally $l$). We set
    $z_H=1$.  The blue, red and yellow curve correspond to $(1-z_*,l)=(10^{-3},2.739),
  (10^{-4},3.461)$ and $(10^{-5},4.173)$, respectively.}
\label{fig:BHE}
\end{minipage}
\hspace{0.5cm}
\begin{minipage}[b]{0.45\linewidth}
\centering
\includegraphics[width=\textwidth]{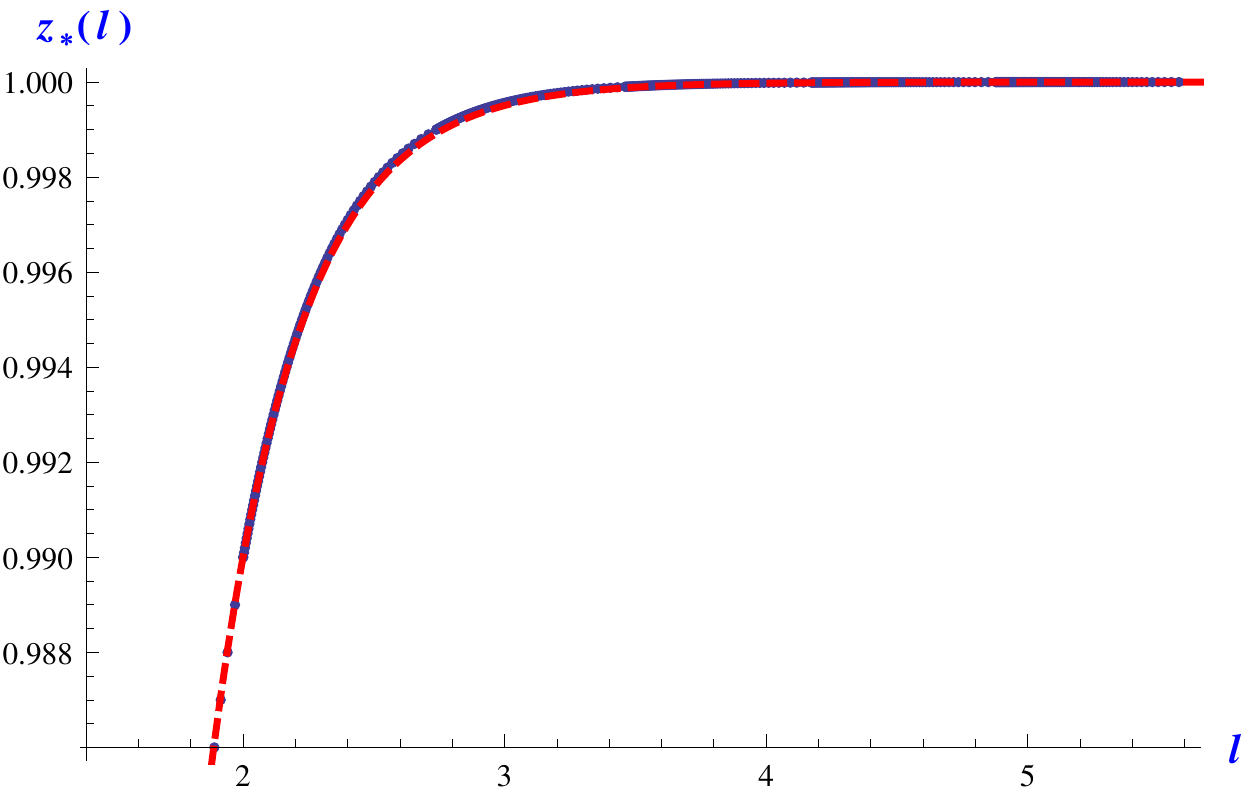}
\caption{Plot of $z_{\star}(l)$ as a function of $l$. Here we have again chosen $z_H=1$.
The blue dots are numerically evaluated points, while the dotted red line
is a best fit to the data with a function of the form $z_H - \alpha e^{-\beta l}$.
The best fit values are $\alpha = 4.106, \ \beta = 3.008 \approx 3$.}
\label{fig:Zl}
\end{minipage}
\end{figure}

\section{HEE in Flat Space Holography}

 In principle, we can study the HEE for the Einstein gravity on a flat space.
As studied from the viewpoint of HEE in \cite{LiTa,NRT}, its dual field theory is
expected to be highly non-local and is not a conventional quantum field theory.

  Consider the four dimensional Minkowski spacetime $R^{1,3}$,
 where the coordinate is written as $(t,x,y,z)$.
We regard $z_b$ as the boundary of the spacetime and assume the total spacetime is simply
given by $z<z_b$. We consider the subsystem $A$ to be a round disk with radius $l$.
The minimal surface is identical to this round disk at $z=z_b$. Therefore
we can show
\ba
\Delta S_A(\omega,k,l)&\propto & \int^l_0 rdr\int^{2\pi}_{0}d\theta e^{ir(k_x\cos\theta+k_y\sin\theta)}e^{-i\omega t}e^{ik_z z},\no
&\propto & \f{l}{k}J_1(kl)e^{-i\omega t+ik_z z_b}.
\ea
where $k_z=\pm \s{\omega^2-k^2}$.

Thus we find that this satisfied the following constraint equation:
\be
\left[\f{\de^2}{\de l^2}-\f{1}{l}\f{\de}{\de l}-\f{\de^2 }{\de x^2}-\f{\de^2 }{\de y^2}\right]\Delta S_A=0. \label{flatee}
\ee
This suggests that the presence of $-3/l^2$ term in (\ref{xx}) corresponds to the negative cosmological constant.
Moreover this agrees with the result (\ref{adsbhem}) for the AdS black hole in the large $l$ limit.

%
%
\providecommand{\href}[2]{#2}\begingroup\raggedright\endgroup

\end{document}